\documentclass[aps,prl,preprintnumbers,amsmath,amssymb,superscriptaddress,twocolumn]{revtex4-2}
\usepackage{color}
\usepackage{float}
\usepackage{subfigure} 
\usepackage{multirow}
\usepackage{booktabs}

\usepackage{graphicx} 

\begin{document}
\newcommand{\bs}[1]{\boldsymbol{#1}}
\newcommand{\dsum}{\displaystyle\sum}
\renewcommand{\.}{\boldsymbol{\cdot}}

\newcommand{\br}{\color{red}}
\newcommand{\bb}{\color{black}}
\newcommand{\revwjq}[1]{\textcolor{black}{#1}}

\title{Optical intensity-gradient torque due to chiral multipole interplay}


\author{Jiquan Wen}
\affiliation{School of Automation, Guangxi University of Science and Technology, Liuzhou, Guangxi 545006, China}
\affiliation{School of Electronic Engineering, Guangxi University of Science and Technology, Liuzhou, Guangxi 545006, China}

\author{Huajin Chen}
\email[]{huajinchen13@fudan.edu.cn}
\affiliation{School of Electronic Engineering, Guangxi University of Science and Technology, Liuzhou, Guangxi 545006, China}
\affiliation{State Key Laboratory of Surface Physics and Department of Physics, Fudan University, Shanghai 200433, China}
\affiliation{Guangxi Key Laboratory of Multidimensional Information Fusion for Intelligent Vehicles, Liuzhou, Guangxi 545006, China}

\author{Hongxia Zheng}
\email[]{hxzheng18@fudan.edu.cn}
\affiliation{School of Electronic Engineering, Guangxi University of Science and Technology, Liuzhou, Guangxi 545006, China}
\affiliation{State Key Laboratory of Surface Physics and Department of Physics, Fudan University, Shanghai 200433, China}
\affiliation{Guangxi Key Laboratory of Multidimensional Information Fusion for Intelligent Vehicles, Liuzhou, Guangxi 545006, China}

\author{Xiaohao Xu}
\email[]{xuxhao\_dakuren@163.com}
\affiliation{State Key Laboratory of Transient Optics and Photonics, Xi'an Institute of Optics and Precision Mechanics, Chinese Academy of Sciences, Xi'an 710119, China}

\author{Shaohui Yan}
\affiliation{State Key Laboratory of Transient Optics and Photonics, Xi'an Institute of Optics and Precision Mechanics, Chinese Academy of Sciences, Xi'an 710119, China}

\author{Baoli Yao}
\affiliation{State Key Laboratory of Transient Optics and Photonics, Xi'an Institute of Optics and Precision Mechanics, Chinese Academy of Sciences, Xi'an 710119, China}

\author{Zhifang Lin}
\affiliation{State Key Laboratory of Surface Physics and Department of Physics, Fudan University, Shanghai 200433, China}
\affiliation{Collaborative Innovation Center of Advanced Microstructures,
Nanjing University,
Nanjing 210093, China}



\begin{abstract}
Owing to the ubiquity and easy-to-shape property of optical intensity, the intensity gradient force of light has been most spectacularly exploited in optical manipulation of small particles. Manifesting the intensity gradient as an optical torque to spin particles is of great fascination on both fundamental and practical sides but remains elusive. Here, we uncover the existence of the optical intensity-gradient torque in the interaction of light with chiral particles. Such a new type of torque derives from the interplay between \revwjq{chirality induced multipoles}, which switches its direction for particles with opposite chirality. We show that this torque can be directly detected by a \revwjq{simple} standing wave field, created with the interference of two counterpropagating plane-like waves. Our work offers a unique route to achieve rotational control of matter by tailoring the field intensity of Maxwell waves. It also establishes a framework that maps a remarkable connection among the optical forces and torques, across chiral to nonchiral.
\end{abstract}


\maketitle

An ongoing endeavor in optomechanics is to achieve accurate, on-demand control of minute particles, for advanced applications in imaging, precise measurements and sensing \cite{ahn2020ultrasensitive,gonzalez2021levitodynamics, millen2020optomechanics,sun2024ai}. Central to this endeavor is to understand optical forces and torques in a way, traceable to the structure of light and particle properties. When light impinges on a particle, optically excited multipoles interact with the excitation field and themselves, giving rise to interception (or extinction) and recoil (or scattering) mechanical effects, respectively \cite{chen2011optical,nieto2010optical,jiang2015universal,zhou2022observation,shi2023advances}. In these interaction processes, the optical force may be produced by the intensity gradient \cite{ashkin1986observation,nieto2010optical}, and other field properties such as the phase gradient \cite{roichman2008optical}, momentum and reactive or imaginary Poynting momentum (IPM) \cite{nieto2010optical,jiang2016decomposition,xu2019azimuthal,zhou2022observation}, which complement the translational control of the particle. Akin to the force, the optical torque is associated with both light extinction and scattering, but the interplay between different-type multipoles does not contribute to the angular momentum transfer \cite{jiang2015universal,wei2022optical}, which prevents a spectrum of field quantities from being coupled to the torque on normal particles. Consequently, the generation of a torque has relied heavily on the optical spin \cite{rashid2018precession,reimann2018ghz,ahn2018optically,jin20216,ahn2020ultrasensitive}, the magnitude of which is limited by the intensity. 

Exploiting particle chirality would open unexpected opportunities for rotational manipulation. This has been demonstrated for dipolar chiral particles, on which the kinetic momentum manifests itself as a torque \cite{bliokh2014magnetoelectric,canaguier2015chiral}, with important applications in enantiomer identification. The main aim of this letter is to ask whether the torque can be induced by the intensity inhomogeneity, which was previously considered responsible only for the force. We demonstrate, analytically and numerically, the existence of such intensity gradient torque for chiral particles, which descends from the recoil effects and exhibits an anti-asymmetry with respect to the electric and magnetic responses of the particle. We also present a general framework for classifying the optical torques according to their field-related properties, in the spirit of the classification of optical forces. 
 
\begin{figure}
    \centering
    \subfigbottomskip=2pt
    \subfigure{
		\label{fig1a}
		\includegraphics[width=\linewidth]{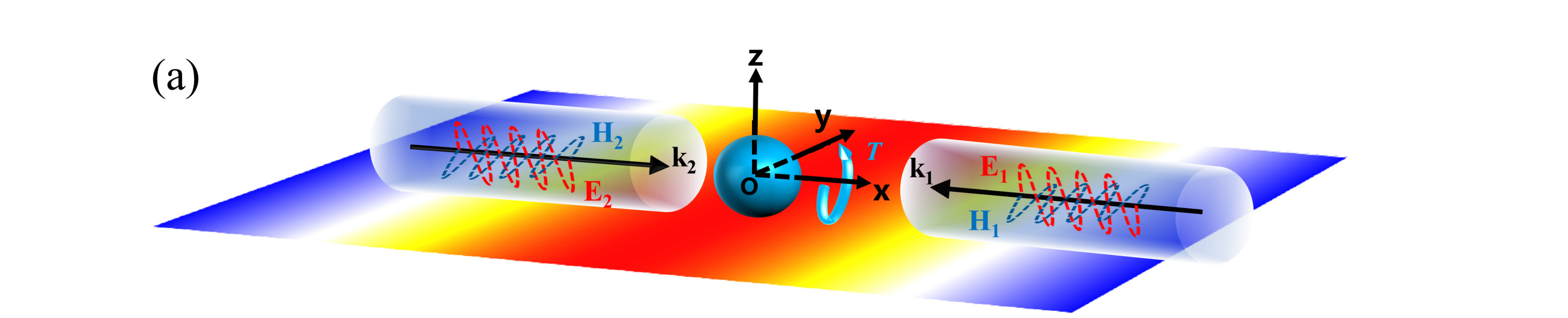}}

    \subfigure{
		\label{fig1b}
		\includegraphics[width=4cm]{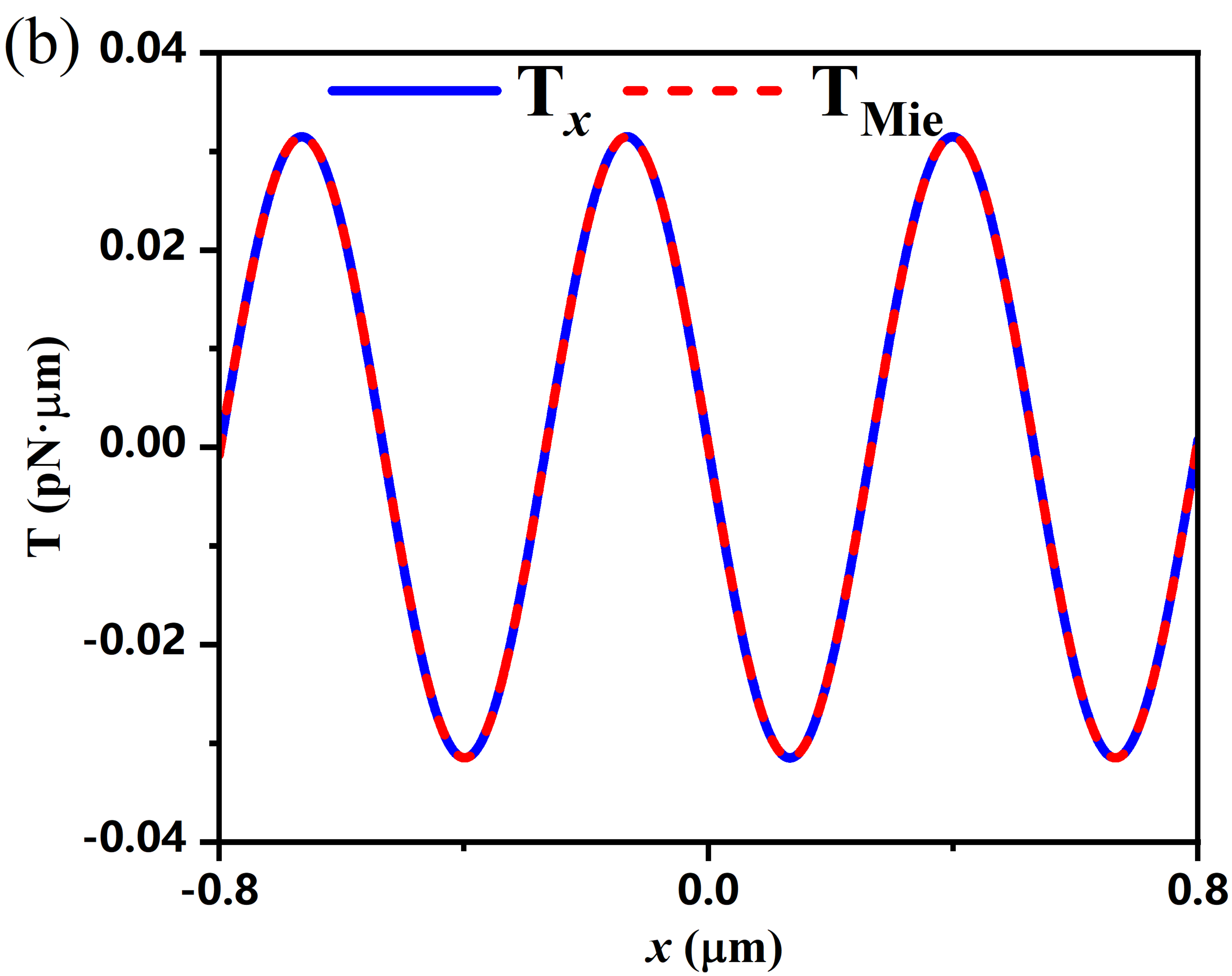}}
  \,
      \subfigure{
		\label{fig1c}
		\includegraphics[width=4cm]{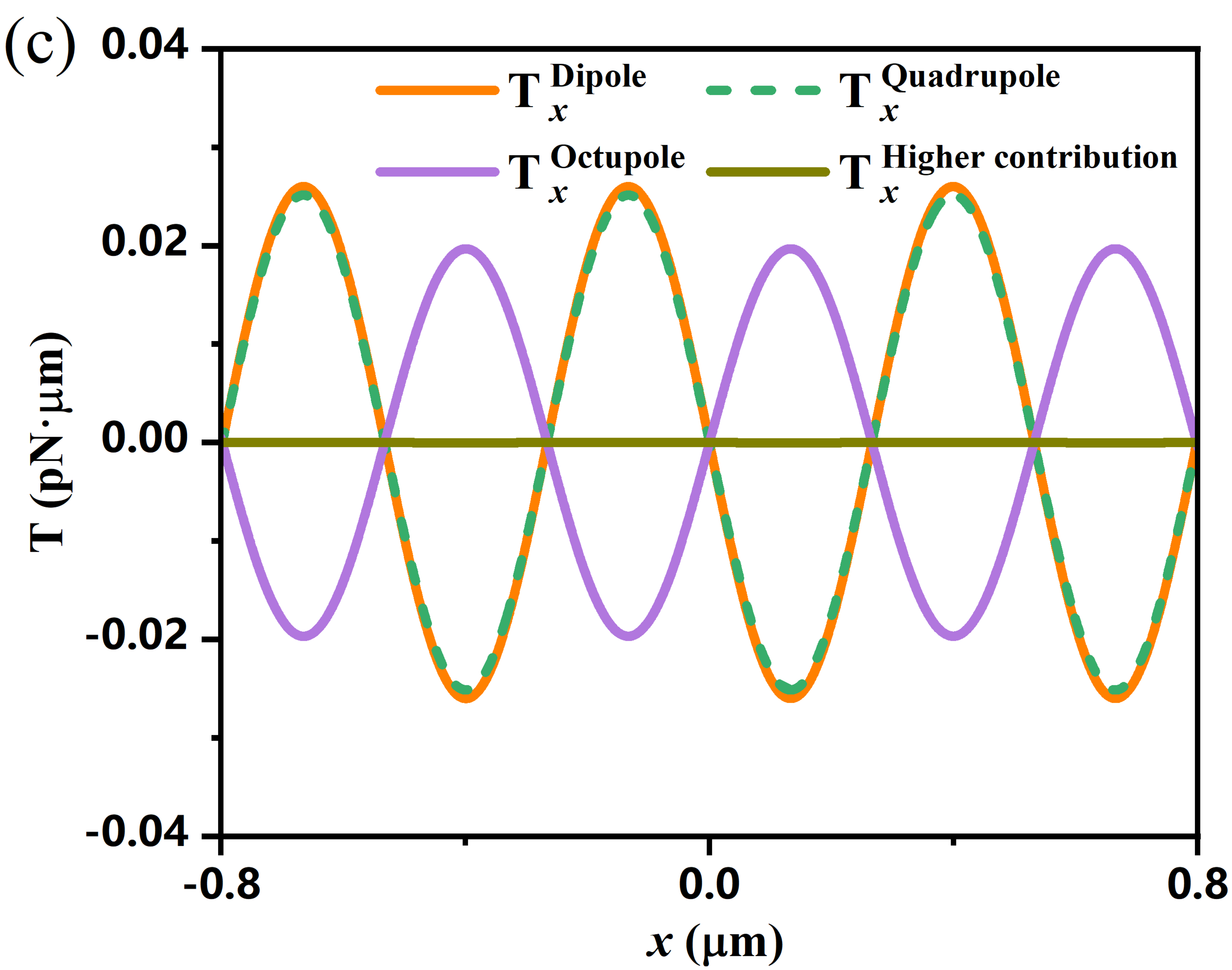}}
\caption{\revwjq{(a)} Schematic illustration of a chiral sphere illuminated by a standing wave composed of two plane waves with identical linear polarization, propagating in opposite directions. The electric field is polarized parallel to the $z$-axis, and the wave vectors $\bf{k}_{\rm 1,2}$ lie on the $xoy$-plane. \revwjq{The} electric intensity gradient is proportional to \revwjq{its} magnetic \revwjq{counterpart} with a negative scale coefficient. \revwjq{(b) Time-averaged optical torque and (c) its multipolar contributions as a function of the particle's position in the Cartesian coordinate system. The particle parameters satisfy $\varepsilon_s=2.5+0.25i$, $\mu_s=1$, $\kappa=0.1$, $r=0.5\mu m$ and the incident wavelength is $\lambda=1.064 \mu m$. 
}\label{fig1}}
\end{figure}

We start by considering a simple inhomogeneous field consisting of two counter-propagating plane waves polarized along $z$ direction, as depicted in Fig. \ref{fig1a}. The electric and magnetic vectors of illumination are given by: 
\begin{equation}
    \mathbf{E}=-2E_0\cos(kx)\mathbf{\hat{z}}, \,\, \mathbf{B}=2\frac{E_0}{c}\sin(kx)\mathbf{\hat{y}}. 
\end{equation}
This standing wave is incident on a sphere made of bi-isotropic chiral material described by the constitutive relation \cite{wang2014lateral}:
\begin{equation}
	\textbf{D}=\varepsilon_0 \varepsilon_s\textbf{E}+ i\kappa \sqrt{\varepsilon_0 \mu_0}\,\textbf{H},\quad	\textbf{B}=\mu_0 \mu_s\textbf{H}-i\kappa \sqrt{\varepsilon_0 \mu_0}\,\textbf{E},
\end{equation}
where the parameter $ \kappa $ describes the particle chirality, and $ \varepsilon_0 (\mu_0) $ and $ \varepsilon_s (\mu_s) $ denote the vacuum and relative permittivity (permeability), respectively. 

\revwjq{In this context, one might not expect the appearance of torque, because the illumination is linearly polarized, carrying no spin angular momentum, and the field momentum is completely canceled by the counter-propagating configuration over the whole space.} However, our calculations based on the Lorenz-Mie method show that the particle 
will experience a torque in the $x$ direction. Also, this torque holds a spatial characteristic consistent with that of the intensity gradient, $\nabla|\mathbf{E}|^2\propto \sin(2kx)$, which reaches zero at the field nodes and antinodes ($kx = n\pi/2$ for integer $n$). 

To reveal the origin of this torque, we resort to the Cartesian multipole expansion model outlined in Ref. \cite{jiang2015universal} and express the torque in terms of the incident field and particle's material properties, by which we arrive at (detailed in Supplemental Material
\cite{SM}):
\begin{align}
    \mathbf{T}&=\sum_{l=1}^N\beta^{(l)}\,\Re\bigl[\gamma^{(l)}_{\text{elec}}\gamma^{(l)\,*}_{\text{x}}-\gamma^{(l)}_{\text{mag}}\gamma^{(l)\,*}_{\text{x}} \bigr]\nabla|\mathbf{E}|^2,\notag\\[2mm]
   \beta^{(l)}&=-\frac{1}{32\,\pi}\frac{(-2)^l (l+1)^2 (l-1)!}{l^2 (2 l-1)! (2 l+1)\text{!!}},
\end{align}
where $ \gamma^{(l)}_{\text{x}} $, $ \gamma^{(l)}_{\text{elec}} $ \revwjq{($ \gamma^{(l)}_{\text{mag}} $)} represent the polarizabilities due to chiral magnetoelectric coupling, \revwjq{and conventional} electric \revwjq{(magnetic)} responses, respectively. The result obtained from this analytical expression (see the blue line in Fig. 1(b)) agrees with the Lorenz-Mie outcome.

Eq. (3) explicitly shows that the torque originates \revwjq{exclusively} in the intensity gradient of illumination. The standing wave thus serves as an ideal platform to \revwjq{experimentally} observe \revwjq{such an intriguing phenomenon}. We remark that Eq. (3) is derived from the recoil part of the total torque with the vanishing interception part \revwjq{\cite{jiang2015universal}}. It follows that the torque is associated with the interplay of multipoles excited in the particle, as reflected by the product terms of polarizabilities in Eq. (3). Further calculations show that the torque on this particle is dominated by the dipole, quadrupole, and octupole responses, while higher-order contributions are negligible. 

\begin{figure}[!ht]
	\includegraphics[width=\linewidth]{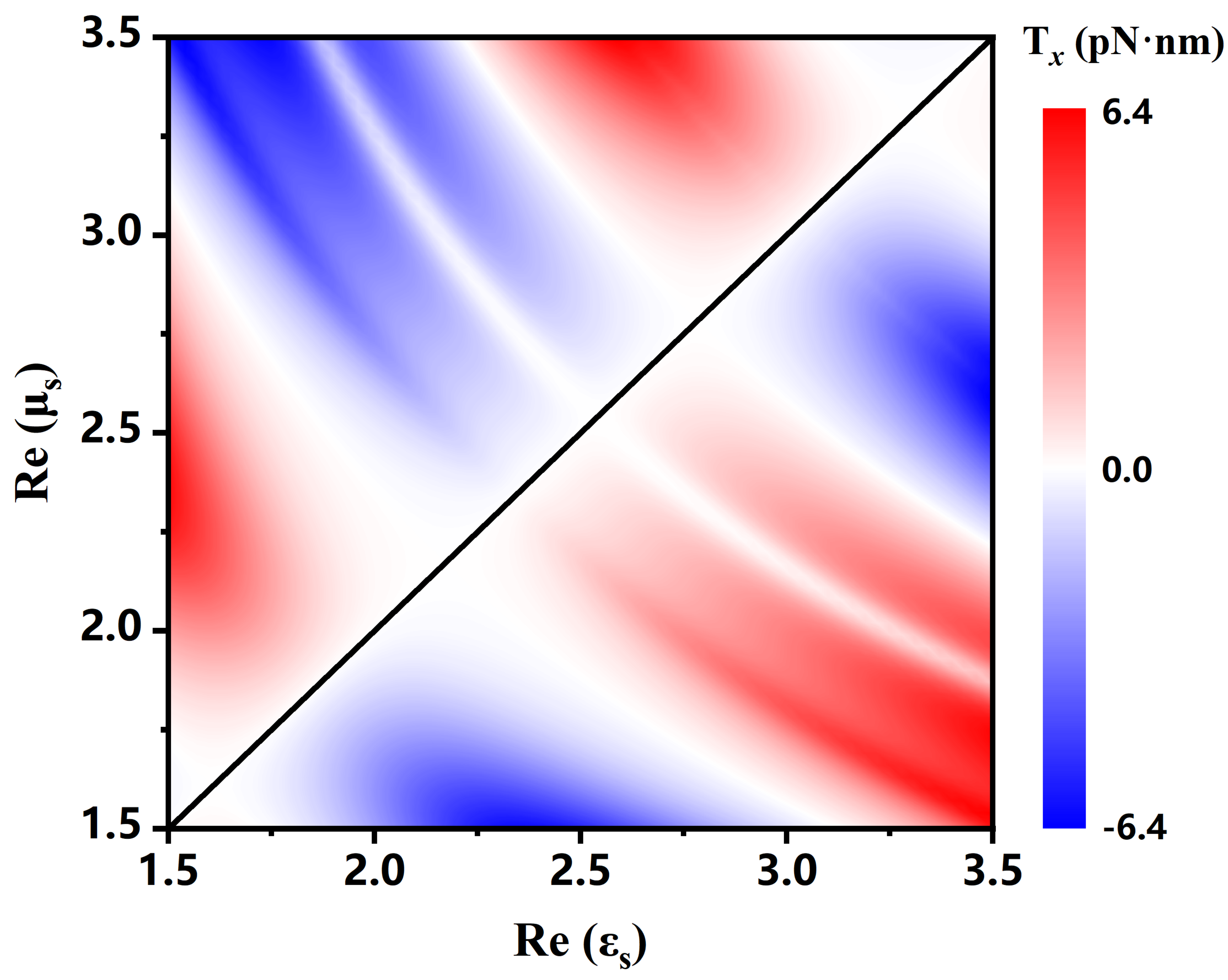}
	\caption{The \revwjq{optical intensity-gradient torque} versus the real parts of the particle's permittivity $ \text{Re}(\varepsilon_s) $ and permeability $ \text{Re}(\mu_s) $ when a particle is immersed in vacuum. Their imaginary parts are fixed at $\Im(\varepsilon_s) = \Im(\mu_s)=0.01 $. }
	\label{fig.fig-3}
\end{figure}

We also notice in Eq. (3) that the intensity gradient torque exhibits an anti-symmetry concerning the exchange of the electric and magnetic polarizabilities: $\gamma^{(l)}_{\text{elec}}\rightleftarrows\gamma^{(l)}_{\text{mag}}$. This property is clearly shown in Fig. \ref{fig.fig-3}, in which the torque is plotted versus the real parts of the particle's relative permittivity $ \text{Re}(\varepsilon_s) $ and permeability $ \text{Re}(\mu_s) $, with $ \Im(\varepsilon_s) = \Im(\mu_s)=0.01 $ and the particle placed at $ (0.3,0,0)\,\mu$m; other parameters are consistent with those in Fig. \ref{fig1a}. Note that $\varepsilon_s$ and $\mu_s$ are linked to the polarizabilities by the same function according to the Mie theory: $\gamma^{(l)}_{\text{elec}} = F(\varepsilon_s,\sqrt{\varepsilon_s\mu_s})$ and $\gamma^{(l)}_{\text{mag}} = F(\mu_s,\sqrt{\varepsilon_s\mu_s})$, so that the exchange $\varepsilon_s\rightleftarrows\mu_s$ gives rise to $\gamma^{(l)}_{\text{elec}}\rightleftarrows\gamma^{(l)}_{\text{mag}}$. As expected, the result is anti-asymmetric with respect to the black line of \(\mu_s = \varepsilon_s\), where the torque achieves zero values. Therefore, the generation of the intensity-gradient torque entails the breaking of \revwjq{electric-magnetic response symmetry (EMRS)}: $\gamma^{(l)}_{\text{elec}} \neq \gamma^{(l)}_{\text{mag}}$ or $\varepsilon_s\neq\mu_s$.
In this regard, this torque resembles the IPM force, for which the breaking of \revwjq{EMRS} is also required \cite{zhou2022observation,chen2020lateral}.

\begin{figure}[!ht]
	\includegraphics[width=\linewidth]{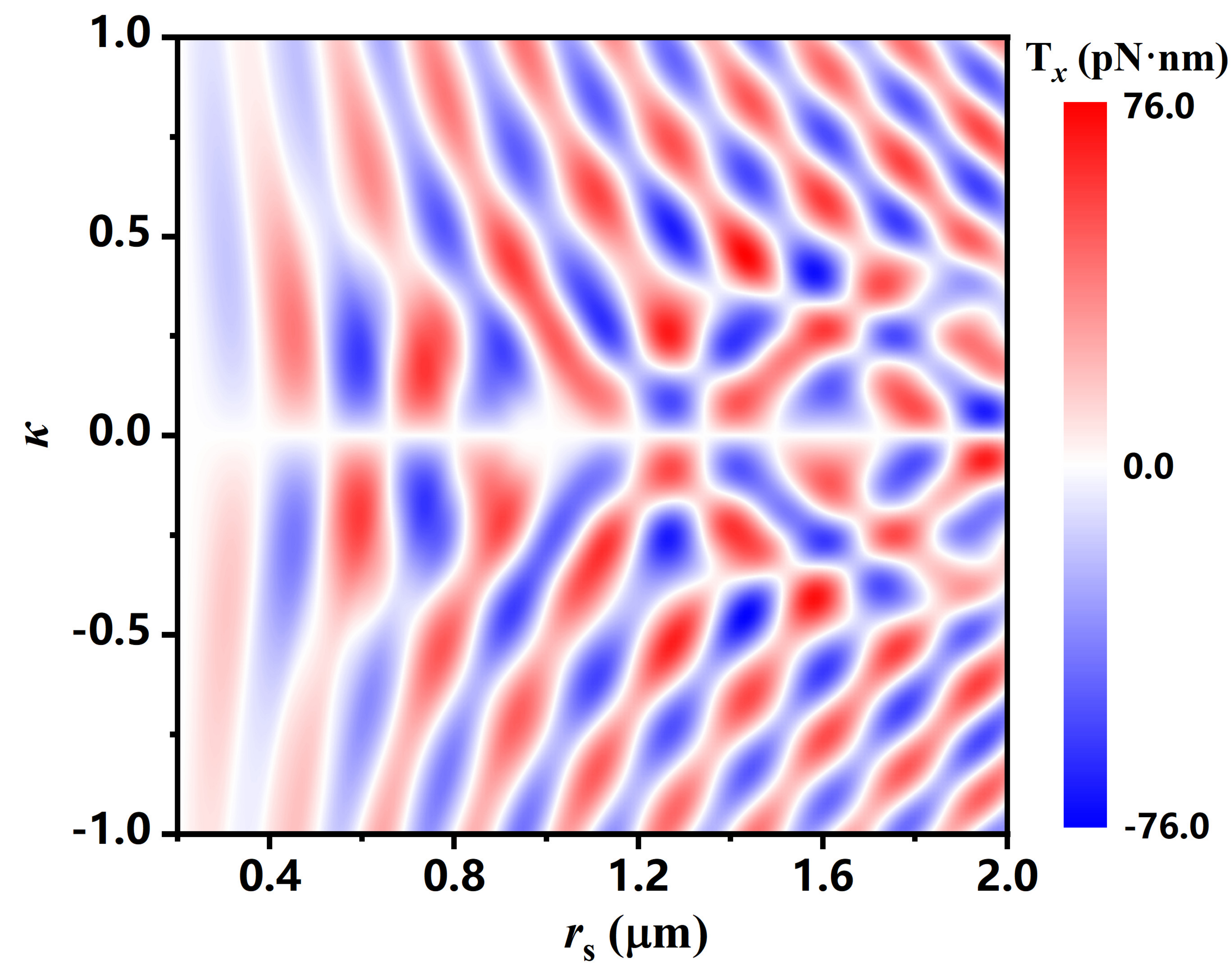}
	\caption{The spatial profiles of the torque versus the chiral parameter $\kappa$ and radius $r_s$. The particles were placed at $ (0.3,0,0)\,\mu m $ in vacuum and the other parameters are consistent with Fig. \ref{fig1b}.}
	\label{fig.fig-4}
\end{figure}

Figure \ref{fig.fig-4} emphasizes the chiral nature of the intensity-gradient torque. Specifically, changing the sign of the particle's chiral parameter \( \kappa \) leads to a direction reversal of the torque experienced by the particle, so that the torque \revwjq{vanishes} for an achiral particle (\(\kappa\) = 0). This chirality-dependent property is explained by the polarizability $\gamma^{(l)}_{\text{x}}$ in Eq. (3), which is an odd function of the chiral parameter: $\gamma^{(l)}_{\text{x}}(\kappa) = -\gamma^{(l)}_{\text{x}}(-\kappa)$, \revwjq{irrespective of particle size \cite{zheng2021optical}}. It is noteworthy that the results depicted in Fig. \ref{fig.fig-4} encompass particle diameters ranging from $0.4\lambda $ to $4\lambda$, substantiating the universality of the analytical formulas Eq. (3) concerning particle size. 

\begin{table*}[ht]
    \centering
    \caption{A list of optical force-torque counterparts, highlighting their relevance to light structure and generation requirements on material properties.}
    \begin{ruledtabular}
    \begin{tabular}{cccccc}
        \multirow{3}{*}{Force/torque categories}& \multirow{3}{*}{Light structure} &\multicolumn{4}{c}{{Material properties (force/torque)}}\\
         \cline{3-6}
        &&Chirality&\revwjq{EMRS} breaking&High-order&Multipole\\
        &&&&multipoles&interplay\\     
        \midrule
        Intensity gradient& $\nabla\left| \mathbf{E} \right|^{2}$ or $\nabla\left| \mathbf{H} \right|^{2}$ & No/Yes &No/Yes&No/Yes &No/Yes\\
        Phase gradient& Im$ \bigl[\mathbf{E}^*\cdot\left(\nabla \mathbf{E}\right)\bigr]$ or Im$ \bigl[\mathbf{H}^*\cdot\left(\nabla \mathbf{H}\right)\bigr]$& No/Yes &No/No&No/Yes &No/No\\
        Kinetic momentum & Re$\left( \mathbf{E}^*\times\mathbf{H}\right)$& No/Yes &No/No&No/No &Yes/No\\
        Reactive momentum & Im$\left( \mathbf{E}^*\times\mathbf{H}\right)$& No/Yes &Yes/ Yes &No/No &Yes/Yes\\        
        Spin & Im$\left( \mathbf{E}^*\times\mathbf{E}\right)$ or Im$\left( \mathbf{H}^*\times\mathbf{H}\right)$& Yes/No &No/No&No/No &No/No\\    
        Reactive helicity gradient &$\nabla$Re$\left(\mathbf{E}^*\cdot\mathbf{H}\right)$& Yes/No &Yes/Yes& No/Yes &Yes/No\\  
        Momentum curl &$\nabla\times$Re$\left(\mathbf{E}^*\times\mathbf{H}\right)$& Yes/No &No/No&No/Yes &No/No\\
    \end{tabular}
    \end{ruledtabular}
\end{table*}

The above results have exemplified the existence of the intensity gradient torque in the special case of standing waves. To generalize this concept, we then assume a generic illumination and incorporate the angular spectrum theory \cite{zhou2022observation} into the Cartesian multipole expansion method \cite{jiang2015universal}. After a lengthy algebra, we decompose the total torque into two parts \cite{SM}:
\begin{equation}
    \mathbf{T} = \sum_{l=1}^{N}\mathbf{T}^{(l)}_{\rm achiral} + \sum_{l=1}^{N}\mathbf{T}^{(l)}_{\rm chiral},
\end{equation}
where $\mathbf{T}_{\rm achiral}^{(l)}$ and $\mathbf{T}_{\rm chiral}^{(l)}$
represent the $2^l$-pole contributions independent and dependent of the particle chirality, respectively. The former was shown associated to the optical spin, the momentum curl and the gradient of reactive helicity \cite{nieto2021reactive} of illumination in a recent work. 
We find that it is constructive to categorize the $l$-order chiral torque into six components according to the field-related quantities:
\begin{equation}
     \mathbf{T}_{\rm chiral}^{(l)} = \mathbf{T}_{\rm spin}^{(l)} + \mathbf{T}_{\rm curl}^{(l)} + \mathbf{T}_{\rm IG}^{(l)} + \mathbf{T}_{\rm PG}^{(l)} + \mathbf{T}_{\rm PM}^{(l)} + \mathbf{T}_{\rm IPM}^{(l)},  \\
 \end{equation}
with
\begin{equation}\label{TC}
\begin{split}
    & \mathbf{T}_{\rm spin}^{(l)} = \hat{\mathcal{C}}_{\rm spin}^{(l)}[\Im(\mathbf{E^*\times E}) + c^2\,\Im(\mathbf{B^*\times B})], \\
     &\mathbf{T}_{\rm curl}^{(l)} = \hat{\mathcal{C}}_{\rm curl}^{(l)}\nabla\times\Re(\mathbf{E^*\times B}), \\
     &\mathbf{T}_{\rm IG}^{(l)} = \hat{\mathcal{C}}_{\rm IG}^{(l)}[\nabla|\mathbf{E}|^2 - c^2\,\nabla|\mathbf{B}|^2], \\
      &\mathbf{T}_{\rm PG}^{(l)} = \hat{\mathcal{C}}_{\rm PG}^{(l)}[\Im(\nabla\mathbf{E}\cdot\mathbf{E}^*)+c^2\,\Im(\nabla\mathbf{B}\cdot\mathbf{B}^*)], \\
       &\mathbf{T}_{\rm PM}^{(l)} = \hat{\mathcal{C}}_{\rm PM}^{(l)}\Re(\mathbf{E^*\times B}), \\
       &\mathbf{T}_{\rm IPM}^{(l)} = \hat{\mathcal{C}}_{\rm IPM}^{(l)}\Im(\mathbf{E^*\times B}), \\
\end{split}
\end{equation}
where the prefactors are determined by the material properties of the particle \cite{SM}. The first two terms, $\mathbf{T}_{\rm spin}^{(l)}$ and $\mathbf{T}_{\rm curl}^{(l)}$, indicate that the optical spin and momentum-curl account for the chiral torque, in addition to the achiral one. 
It is interesting to note that the optical spin is coupled to the chiral torque by its dual-symmetric form \cite{bliokh2013dual,golat2024electromagnetic}, $\Im(\mathbf{E^*\times E}) + c^2\,\Im(\mathbf{B^*\times B})$, but the electric and magnetic spin contribute differently to the achiral torque in general \cite{wei2022optical}. 

The field quantities arising in the last four components are unavailable to the achiral torque, 
but they are responsible for the well-known intensity-gradient force \cite{ashkin1986observation,nieto2010optical}, phase-gradient force \cite{roichman2008optical}, radiation pressure and IPM force on achiral particles \cite{nieto2010optical,jiang2016decomposition,zhou2022observation}. In this connection, $\mathbf{T}_{\rm IG}^{(l)}$, $\mathbf{T}_{\rm PG}^{(l)}$, $\mathbf{T}_{\rm PM}^{(l)}$ and $\mathbf{T}_{\rm IPM}^{(l)}$ represent the angular analogs of these types of force. It is worth noting that the standard momentum, $\Re(\mathbf{E^*\times H})$, is sometimes written as the sum of its spin and orbital parts, especially in literature focusing on the optical force due to Belinfante's spin momentum (BSM) \cite{bliokh2014extraordinary,bekshaev2015transverse,zhou2023optical,yu2023anomalous}. One may also perform the spin-orbit decomposition for the momentum in $\mathbf{T}_{\rm PM}^{(l)}$ so that the BSM can manifest itself as the torque, but we would keep its original form for the mathematical elegance of our theory. We stress that the IPM torque $\mathbf{T}_{\rm IPM}^{(l)}$, akin to the IPM force \cite{nieto2010optical,zhou2022observation}, can act on a dipole particle for a generic field, whereas the intensity gradient torque, $\mathbf{T}_{\rm IG}^{(l)}$, requires the excitation of high-order ($l > 1$) multipoles, as $\hat{\mathcal{C}}_{\rm IG}^{(l)} = 0$ for $l = 1$. However, in some special fields (e.g., the standing wave shown in Fig. 1a), the reactive momentum, $\Im(\mathbf{E^*\times B})$, may reduce to its orbital part \cite{nieto2021reactive,nieto2022complex} (which is proportional to $\nabla|\mathbf{E}|^2-c^2\nabla|\mathbf{B}|^2$), such that the intensity gradient induces the torque via the dipole response (Fig. 1c).

To gain an overview of the principle of optical force and torque, Table 1 compares the torques with their force counterparts \cite{nieto2010optical,zhou2022observation,wang2014lateral,hayat2015lateral,zheng2020general} featuring the same light structure, in terms of the generation requirements on material properties. It is of marked interest that each type of optical force (or torque) for an achiral particle finds its analog in the torque (or force) on a chiral one. The breaking of \revwjq{EMRS} \cite{chen2020lateral} is necessary for the intensity-gradient torque and all mechanical effects induced by reactive quantities. High-order multipole excitation is a requisite for torques, except for the conventional spin torque and those due to the field momenta including kinetic and reactive. Also, the mechanical manifestations of some field quantities (e.g., the IPM force and the intensity gradient torque) must rely on the multipole interplay, which highlights their recoil nature. At last, we remark that the gradient of standard helicity (not listed in the table), $\nabla\Im(\mathbf{E}^*\cdot\mathbf{H})$, may also contribute to the optical force for chiral particles \cite{wang2014lateral,zheng2020general}, but is shown irrelevant to the torque in light of our theory.

In summary, we have discovered the intensity gradient torque by exploiting the interplay of chiral multipoles, and have built a generic multipole model, which expands the classifying framework of optical torques from nonchiral 
to chiral. The intensity gradient torque offers unique opportunities for particle spinning in terms of frequency and degree of freedom, in that the intensity inhomogeneity may be significant with a small power density \cite{ashkin1986observation}, and in that the distribution of optical intensity, in practice, is easier to customize than phase and polarization. It also opens new possibilities for near-field tweezers \cite{ju2023near,zhang2021plasmonic}, in which the field localized beyond the diffraction limit is expected to enhance strongly the torque. In fact, the intensity gradient and spin are compatible properties of light, which promises their integration into chiral and rotational optomechanics. 

As a closing remark, we would point out that material non-reciprocity and anisotropy, which are beyond the scope of this paper, also play an important role in light-matter interaction. Although this subject has recently been of active experimental and numerical research in optical manipulation \cite{bliokh2014magnetoelectric,zhu2018self,li2020momentum,shi2022inverse,nan2023creating,hu2023structured,achouri2023multipolar,riccardi2023electromagnetic}, its theory is yet to be well developed, especially in scenarios of high-order multipoles. We anticipate that our multipole framework, which is established from the fundamental constitutive relation and first principle \cite{SM}, will set \revwjq{methodologically in the framework of classical electromagnetic theory} an inspiring example for the development of optical force and torque theories for non-reciprocal and anisotropic particles.

\begin{acknowledgments}
This work is supported by National Natural Science Foundation of China (No. 12174076, No. 12204117, and No. 12274181), Guangxi Science and Technology Proiect (No. 2023GXNSFFA026002, No. 2024GXNSFBA010261, No. 2021GXNSFDA196001, and No. AD23026117), and the open project of State Key Laboratory of Surface Physics in Fudan University (No. KF2022\_15).
\end{acknowledgments}


\bibliography{ref}

\end{document}